# The Radcliffe Wave is Oscillating

Ralf Konietzka[1,2,3✉], Alyssa A. Goodman[1], Catherine Zucker[1,4], Andreas Burkert[3,5], João Alves[6], Michael Foley[1], Cameren Swiggum[6], Maria Koller[6], Núria Miret-Roig[6]

[1] Harvard University Department of Astronomy, Center for Astrophysics | Harvard & Smithsonian, 60 Garden St. Cambridge, MA 02138, USA
[2] Ludwig-Maximilians-Universität München, Geschwister-Scholl Platz 1, 80539 Munich, Germany
[3] Max Planck Institute for Extraterrestrial Physics, Giessenbachstraße 1, 85748 Garching, Germany
[4] Space Telescope Science Institute, 3700 San Martin Drive, Baltimore, MD 21218, USA
[5] University Observatory Munich, Scheinerstrasse 1, 81679 Munich, Germany
[6] University of Vienna, Department of Astrophysics, Türkenschanzstraße 17, 1180 Vienna, Austria
✉ e-mail: ralf.konietzka@cfa.harvard.edu

**Our Sun lies within 300 pc of the 2.7-kpc-long sinusoidal chain of dense gas clouds known as the Radcliffe Wave.[1] The structure's wave-like shape was discovered using 3D dust mapping, but initial kinematic searches for oscillatory motion were inconclusive.[2–7] Here we present evidence that the Radcliffe Wave is oscillating through the Galactic plane while also drifting radially away from the Galactic Center. We use measurements of line-of-sight velocity[8] for $^{12}$CO and 3D velocities of young stellar clusters to show that the most massive star-forming regions spatially associated with the Radcliffe Wave (including Orion, Cepheus, North America, and Cygnus X) move as if they are part of an oscillating wave driven by the gravitational acceleration of the Galactic potential. By treating the Radcliffe Wave as a coherently oscillating structure, we can derive its motion independently of the local Galactic mass distribution, and directly measure local properties of the Galactic potential as well as the Sun's vertical oscillation period. In addition, the measured drift of the Radcliffe Wave radially outward from the Galactic Center suggests that the cluster whose supernovae ultimately created today's expanding Local Bubble[9] may have been born in the Radcliffe Wave.**

In Figure 1a (Supplementary Figure 1), we present a 3D map showing the most massive star-forming regions (including Orion, Cepheus, North America, and Cygnus X) and embedded young stellar clusters associated with the Radcliffe Wave.[1] As expected, the young clusters and star-forming molecular clouds are co-located in three dimensions. The clusters were selected from a top-down view of the Galaxy, without an *a priori* selection criterion for height off the Galactic plane (see Methods), suggesting that the clusters' stars were born in the molecular clouds along the Radcliffe Wave. So, these clusters' motions can serve as a tracer of the Wave's kinematics. Using clusters' motions rather than individual stars allows for more precise conclusions about kinematics, as individual stars' motions within a host cluster are averaged out, reducing uncertainty.

The kinematic counterpart of the Radcliffe Wave's spatial undulation is displayed in Figure 1b (Supplementary Figure 1), where the *z*-axis shows the vertical velocity of the stellar clusters, after accounting for the Sun's motion.[10] Based on the sinusoid-like shape of the Radcliffe Wave, one could imagine it as either a traveling or a standing wave. Extended Data Figure 1 (and the online version in Supplementary Figure 4) shows how traveling and standing versions of the Radcliffe Wave would appear. For a traveling wave, the wave's spatial shape fully determines vertical velocities, assuming a gravitational potential,[11–13] as explained in the Methods section. For a standing wave, regions located at the zero-crossings of the wave are at rest, and the velocities of regions located at spatial extrema are not constrained by the wave's spatial structure. The fit shown as a solid black line in Figure 1 (cf. Supplementary Figure 1, Methods) indicates that the Radcliffe Wave oscillates through the Galactic plane like a *traveling* wave, such that regions currently at the zero points of the Wave (near the Galactic plane) move through at their greatest vertical velocity, while molecular clouds located at spatial extrema (farthest from the plane) are at their turning points, with zero vertical velocity. While a traveling wave provides an excellent fit to the observations, with a 90° phase offset between position ($z$) and velocity ($v_z$), as evident in Figure 1 (and in the interactive version in Supplementary Figure 1), it also provides an extremum for the wave motion. When fitting a mixture model that allows for superpositions of a traveling and a standing wave, we find that a traveling wave is strongly preferred over a standing wave (see Methods). A standing wave model would have a zero phase offset between positional ($z$) and velocity ($v_z$) variations, which is inconsistent with the data (see Methods).

For stars not far off the Galaxy's disk compared with its scale height, a classical harmonic oscillator is often used to describe stars' responses to the vertical Milky Way potential.[14] By analogy to a pendulum, however, when the amplitude of oscillation is large, one needs to consider non-linear effects that are not accounted for in the formalism describing a purely harmonic oscillator. In the case of the Galaxy, these non-linearities arise from the vertical decrease in the midplane density, becoming significant at vertical positions on the order of the Galaxy's vertical scale height. So, in describing the motion of a pendulum oscillating far from vertical, we model the Wave's motion within the Galactic potential[11–13] as an anharmonic oscillator. In the modeling, we account for the Galaxy's midplane density declining with height and also with galactocentric radius (see Methods). Treating the Radcliffe Wave as a single coherent structure in space and velocity, responding to the Galactic potential, the structure is well-modeled as a damped sinusoidal wave with a maximum amplitude of ~220 pc and a mean wavelength of ~2 kpc. The corresponding maximum vertical velocity is ~14 km s$^{-1}$.

As described in the Methods section, the transparency of the blue points in Figure 1 are coded by a fitting procedure in which statistical inliers (opaque) and outliers (transparent) were explicitly taken into account. Remarkably, we find that the most of the outliers belong to the Perseus and Taurus molecular clouds, which lie on the surface of feedback bubbles.[9,15] The expansion motions of those bubbles likely overwhelm the kinematic imprint of the Wave in the present day.

A schematic view of the vertical oscillation of the stellar clusters in the Radcliffe Wave is shown in Figure 2 (and in the animated view in Supplementary Figure 6). In the static version of Figure 2, we show the excursion of the Wave at the present day along with the best fit at phases corresponding to the minimum (60°) and maximum (240°) deflection of the Wave above and below the Galactic disk. Given the traveling wave nature of the Radcliffe Wave, the animated version of Figure 2 shows that the extrema appear to move from right to left as seen from the Sun. Using the modeled amplitude and wavelength at each end of the Wave, we can calculate the so-called phase velocity describing this apparent motion of the Wave's extrema (see Methods). We find that the phase velocity changes from ~40 km s$^{-1}$ (near CMa OB1) to ~5 km s$^{-1}$ (near Cygnus X).

In addition to its vertical oscillation, we find evidence that the Radcliffe Wave is drifting radially outward from the Galactic Center with a velocity of 5 km s$^{-1}$. This drift occurs in a co-rotating reference frame dictated by Galactic rotation, modeled by the same Milky Way potentials[11–13] used to model the Wave's oscillation. In Extended Data Figure 2b (and in the interactive version in Supplementary Figure 2) we show the vectors representing the Wave's solid body drift in the Galactic plane. The direction of the drift bolsters the previously proposed idea[9] that the Radcliffe Wave served as the birthplace for the Upper Centaurus Lupus and Lower Centaurus Crux star clusters, home to the supernovae that generated the Local Bubble about 15 Myr ago.[9] An exact traceback of the Wave's position over the ~15 Myr since the birth of the Local Bubble would require modeling the deceleration of the Wave as its dense clouds move through a lower-density ISM, which is beyond the scope of this work.

We confirm the motion found in stellar clusters' 3D velocities using $^{12}$CO observations of dense clouds along the Wave, from which we can measure line-of-sight velocities.[8] Since the Radcliffe Wave (length 2.7 kpc) is so close to us (0.25 kpc at the closest point), the lines of sight from the Sun to various clouds in the Wave are oriented at a wide variety of angles relative to an *x, y, z* heliocentric coordinate system, as shown in Extended Data Figure 3b (and in the interactive version in Supplementary Figure 3). Given the Radcliffe Wave's vertical excursions above and below the Sun's position, even purely vertical cloud velocities are probed by a line-of-sight velocity measurement. So, if we treat the Wave as a kinematically coherent structure, we can study 3D motion of its gas just by investigating observed "1-D" radial (LSR) velocities. For more details on the velocity modeling, see the Methods section.

Extant observations combined with our modeling can constrain possible formation mechanisms for the Radcliffe Wave. A gravitational interaction with a perturber seems a natural possible origin,[16] but the Wave's stellar velocities are not fully consistent with models of a perturber-based scenario.[3] In particular, recent studies[3,17] suggest that the dominant wavelength resulting from such a perturbation is an order of magnitude larger than the pattern we observe, challenging this scenario for the Wave's origin. Gas streamers falling onto the disk[18] potentially lead to shorter wavelengths, but modeling has not yet been done to see if this inflowing gas could oscillate on scales commensurate with the Radcliffe Wave. Internal to the disk, a hydrodynamic instability[19] may be able to generate waves on the right scale, but additional work is needed to determine whether such an instability could push gas ~220 pc above the disk, and/or produce a traveling wave. A superposition of feedback-driven structures could reproduce the observed wavelength and amplitude of the Wave, but might require (too much) fine-tuning to also explain the Wave's traveling nature and order-of-magnitude change in phase-velocity. In support of feedback-driven scenarios, we note that relevant galaxy models *do* show nearly straight filaments reminiscent of the Wave (as viewed from the top down) drifting radially[20] in a fashion similar to the observed motion of the Radcliffe Wave. Analyses of future astrometric and spectroscopic measurements of young stars and high-resolution imaging of external galaxies in concert with improved hydrodynamic simulations of galactic-scale features should be able to discriminate amongst potential formation scenarios.

The spatially and kinematically coherent Radcliffe Wave serves as a unique environment for obtaining insights into local Galactic dynamics. As explained above, the kinematic and spatial signature of the Radcliffe Wave can be described in a self-consistent manner, using its response to the local Galactic potential. Near the

Sun, this potential is dominated by the gravitational attraction of baryonic matter in the form of stars and gas.[21] In addition, dark matter, which is known to envelope galaxies in a spherical "dark halo" is also expected to effect the local disk kinematics.[22] In the Methods section, we use the coherent motion of the Radcliffe Wave, assuming it is controlled by the combined gravity of baryonic and dark matter, to estimate the properties of the local Galactic potential. We find a total midplane density of $0.085^{+0.021}_{-0.017}$ M☉pc$^{-3}$, consistent with conventional approaches to deriving properties of the local Galactic mass distribution.[21,23–26] By incorporating direct observations of the gaseous and stellar mass components,[21] and thereby constraining the amount of the baryonic mass component in the solar neighborhood, we also infer the local amount of dark matter (see Methods). Our analysis yields a local density of $0.001^{+0.024}_{-0.021}$ M☉pc$^{-3}$ for a spherical dark halo. If dissipative dark matter exists, it is possible that dark matter may also accumulate in the form of a very thin disk relative to the baryonic mass component.[27] We find an upper bound for a hypothetical dark disk of $0.1^{+3.3}_{-3.3}$ M☉pc$^{-2}$. Simultaneously we also utilize the Radcliffe Wave to derive the frequency of the Sun's oscillation through the Galactic plane (see Methods). We find that the Sun oscillates through the disk of the Milky Way with a period of $95^{+12}_{-10}$ Myr, implying that our Solar system crosses the Galactic disk every $48^{+6}_{-5}$ Myr, consistent with standard values[14] (see Methods).

The measurements of the oscillation and drift of the Radcliffe Wave presented here offer new constraints on the formation of dense star-forming structures within the Milky Way, and the possible origins of kpc-scale wave-like features in galaxies. Upcoming deep and wide surveys of stars, dust and gas will likely uncover more wave-like structures, and measurements of their motions should provide insights into the star formation histories and gravitational potentials of galaxies.

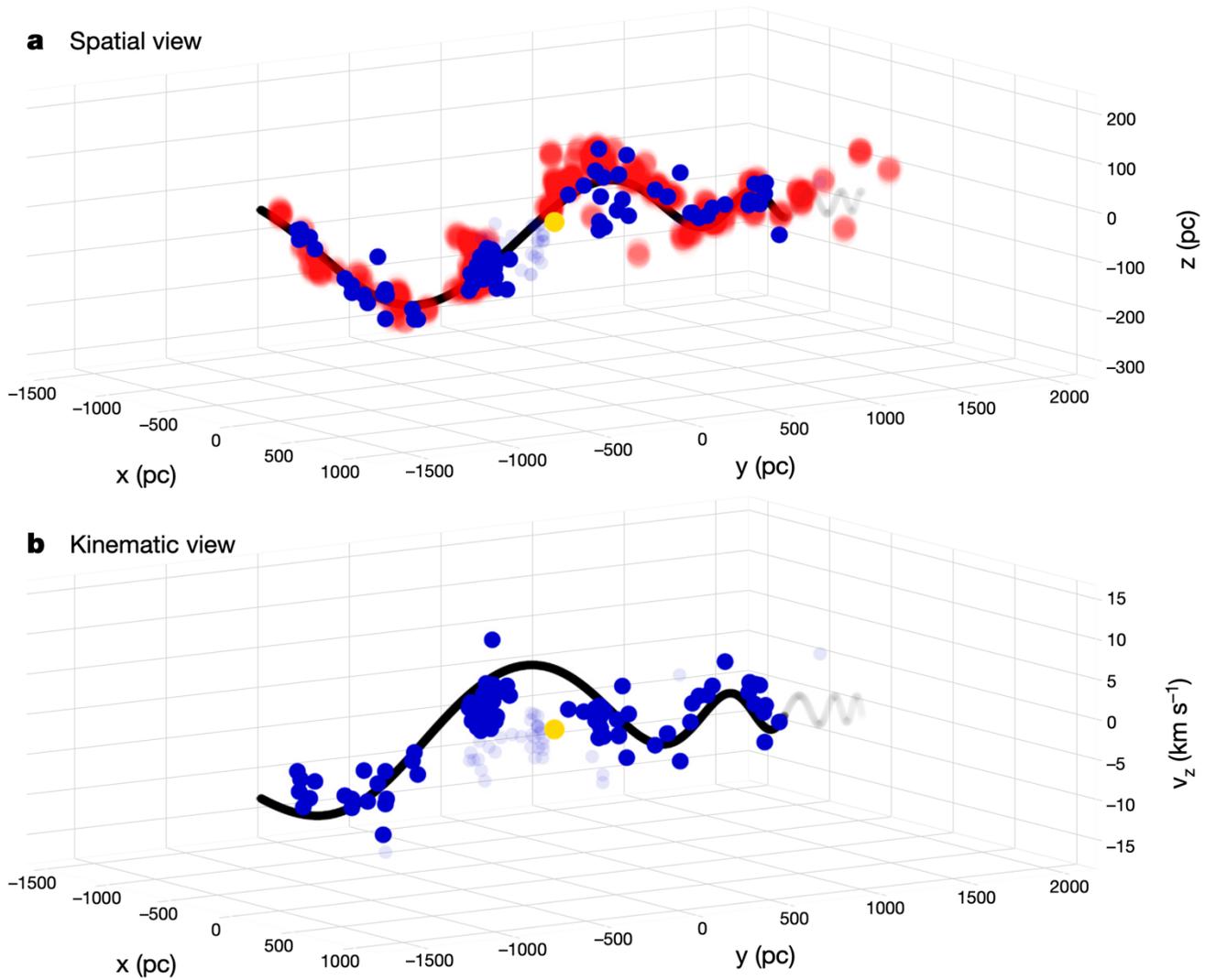

**Figure 1 (Interactive): A spatial and kinematic view of the Solar neighborhood.** *For the best experience, please view the online interactive version in Supplementary Figure 1 (only oscillation) and Supplementary Figure 5 (total motion, combining oscillation + in-plane drift). The variation of the Wave with phase / time in the online interactive versions is obtained by evolving the best fit with phase / time (see Methods), while keeping the distances of the data to the fit constant. On timescales of several tens of Myr, the Radcliffe Wave is likely to be affected by internal (molecular cloud destruction[28]) and global (kinematic phase mixing) effects. As these effects may be superimposed on the Wave's motion in the future, the molecular cloud and star cluster data in the interactive Supplementary Figure 5 are fading out over time.* **Panel a**: *A 3D SPATIAL view of the Solar neighborhood in Galactic Cartesian coordinates (position-position-position space: x, y, z). We show the most massive local star-forming regions spatially associated with the Radcliffe Wave in red. Since we identify the Perseus and Taurus molecular clouds as outliers (see Methods), we do not show these clouds here. The young stellar clusters are shown in blue (inliers opaque / outliers transparent), the Sun in yellow, and the best-fit-model in black. Since we have no star cluster measurements at the right end of the Wave and the amplitude of the ripples in this region approaches the error of the dust distance measurements,[29,30] this section is shown in gray, implying that ripples and no ripples are indistinguishable at this end of the Wave.* **Panel b**: *A 3D KINEMATIC view of the Solar neighborhood (position-position-velocity space: x, y, vertical velocity $v_z$). The colors and symbols are the same as in Panel a.* **The interactive version offers views from any direction.**

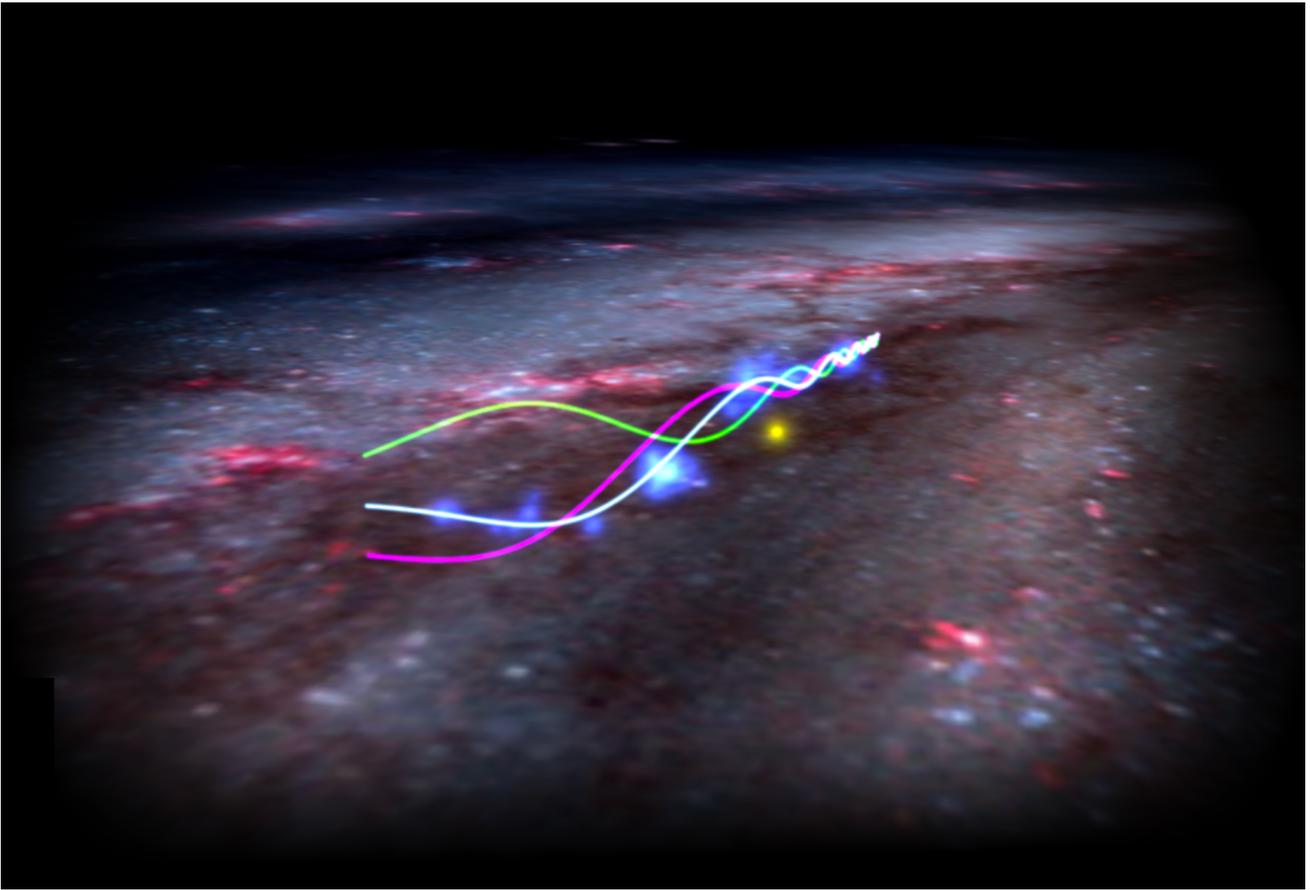

**Figure 2 *(Interactive): A view of the Radcliffe Wave and its oscillatory pattern.*** *The light blue curve shows the traveling wave model presented here, while the blue fuzzy dots show the current positions of the stellar clusters. The magenta and green traces show the Wave's minimum and maximum excursions above and below the plane of the Milky Way, separated by 180° in phase. For more phase snapshots see Extended Data Figure 1, which also compares the motion of the traveling wave to predictions for a standing wave. The Sun is shown in yellow. The background image is an artist's conception of the Solar neighborhood, as seen in WorldWide Telescope. An interactive figure corresponding to this static view is available online as [Supplementary Figure 6](Supplementary Figure 6).*

# Methods

## Stellar Cluster Catalog

The basis of the stellar cluster analysis comprises a large sample size catalog[31] that uses *Gaia*[32] data, omitting lower-quality parallaxes, proper motions, and radial velocities. In addition, we incorporate clusters from other works[33–37] that rely on *Gaia* data, ensuring that only non-duplicate clusters were included in the final catalog. We also create a stellar member catalog, which encompasses all known stars within the identified clusters. We performed cross-matching between the stellar members catalog and the latest *Gaia* data release (*Gaia* DR3[38]). To enhance the number of radial velocity measurements in our final sample, we also make use of the *APOGEE* survey.[39] Finally, for each cluster, we compute the mean position ($x, y, z$) and velocity ($u, v, w$) as well as the standard error of the mean, removing clusters with kinematic anomalies (radial LSR velocities above 50 km s$^{-1}$, typically associated with clusters with a low number of radial velocity measurements). We adopt a peculiar Solar motion[10] of ($u_\odot, v_\odot, w_\odot$) = (10.0, 15.4, 7.8) km s$^{-1}$ and correct the ($u, v, w$) values of each stellar cluster for this Solar motion to obtain its current 3D space velocity with respect to the local standard of rest (LSR) frame ($v_x, v_y, v_z$).

For all clusters, we use the ages as provided in the literature.[31,33–37] We restrict our analysis to only the youngest clusters (ages smaller than 30 Myr), guaranteeing that these associations are still tracing the velocities of the molecular clouds that served as their birthplaces. We make a conservative selection on the *x-y* plane by enclosing all stellar clusters within ±5 times the radius of the Wave (47 pc) from the best fit of the dust component (see "Modeling the Spatial Molecular Cloud Distribution"). We find that the Lacerta OB1 association is the only complex that is vertically (in the Heliocentric Galactic z-direction) not associated with the cold molecular component of the Radcliffe Wave. In addition, the Lacerta OB1 complex is kinematically drifting towards the Solar system, potentially connected to the Cepheus spur,[40] which lies behind the Wave with respect to the Sun. This motion is distinct from the motion of the Radcliffe Wave, which shows evidence of drifting coherently away from the Sun. As this implies that Lacerta is the only stellar cluster complex being spatially and kinematically distinct from the Radcliffe Wave, we do not include its clusters in the final analysis. We compare our cluster sample to a recent open cluster catalog,[41] restricting the analysis again to only the youngest clusters, finding no effect on our results.

## Full Spatial and Kinematic Model

In our analysis of the Radcliffe Wave's kinematics we distinguish between two velocity components with respect to the Galactic plane: 1) "in-plane" motion; and 2) vertical "z" motion. The decoupling of the in-plane and z-motion of the Radcliffe Wave is satisfied by the fact that the maximal vertical deflection of regions along the Wave is at most of the order of the scale height of the Galactic disk.[14] We also examined the effect of a misalignment between the orientation of our adopted Galactic "disk" (parallel to $z = 0$ pc) and the "true" Galactic disk, which could be slightly tilted with respect to the Galactic Cartesian coordinate frame, finding it has having no effect on our results.

For the first component, we take into account Galactic and differential rotation[11–13] and allow the Wave to have an additional overall 2D velocity component, meaning that we permit the structure to move as a solid body through the Galactic plane with a fixed *x*- and *y*-velocity ($v_{x,plane}$, $v_{y,plane}$).

In the case of the *z*-component, we find that the Radcliffe Wave oscillates like a traveling wave driven by the Milky Way's gravitational acceleration. Adopting a typical density ratio of $10^4$ between the Wave and the surrounding more tenuous medium (based on 3D dust mapping[42]), we can assume that the gravitational acceleration induced by the Galactic disk dominates over the gravitational deceleration caused by the surrounding medium. We therefore model the forces acting on the Wave solely by gravity. Since the Radcliffe Wave's amplitude is on the order of the Galaxy's vertical scale height, the vertical gravitational potential $\Phi$ resulting in this acceleration can be represented by an anharmonic oscillator.[43] To capture the anharmonic nature of the oscillation requires that we describe the height variation in the potential up to a fourth order in *z* (as in equation (1), below). To ensure that the potential is symmetric with respect to the Galactic disk we set the prefactors of the first and third order equal to zero. We assume the Galactic density distribution to be axisymmetric and model the dependence of the density on the Galactocentric radius as a decaying exponential function.[22] As the influence of the radial force term in Poisson's equation can be neglected in our region of interest[14,22] (7 to 10 kpc in Galactocentric radius and up to 300 pc in vertical displacement from the Galactic plane) the potential can be expressed as

$$\Phi(z,r) = \frac{1}{2} \omega_0^2 \exp\left((r_\odot - r)\, r_0^{-1}\right) \left(z^2 - \mu_0^2 z^4\right) \tag{1}$$

where r is the Galactocentric radius and *z* the Galactic Cartesian *z* coordinate. We set the distance of the Sun to the Galactic Center, $r_\odot$ equal to 8.4 kpc, according to our adopted Galactic potential model.[11] Note that the distance from the Sun to the Galactic Center can always be absorbed in $\omega_0$ and $r_0$, the zeroth order oscillation frequency and the radial scale length of the potential, respectively. In our region of interest, $\mu_0$ is proportional to the inverse scale height of the potential. Since the scale height can be considered independent of the Galactocentric radius for small variations in radius,[22,44] we can assume $\mu_0$ to be constant in our region of interest. Our model of the potential has therefore in total three free parameters: $\omega_0$, $\mu_0$, and $r_0$. These parameters are adopted by comparing equation (1) with an existing model of the Galactic potential[45] including its updated parameters[11] using a standard chi-square minimization in the range of Galactocentric radius varying from 7 to 10 kpc with a step size of 120 pc and vertical height above and below the Galactic plane varying from 0 to 300 pc with a step size of 12 pc. This leads to $\omega_0$ = 74 km s$^{-1}$ kpc$^{-1}$, $r_0$ = 3.7 kpc and $\mu_0$ = 1.3 kpc$^{-1}$. The Oort constants $A_{Oort}$ and $B_{Oort}$ for the adopted Galactic potential[11] are 15.1 km s$^{-1}$ kpc$^{-1}$ and -13.7 km s$^{-1}$ kpc$^{-1}$. The circular velocity at the position of the Sun is 242 km s$^{-1}$. From our adopted values of $\omega_0$ and $\mu_0$ we infer a local matter density $\rho_0$ = 0.1 M$_\odot$pc$^{-3}$ and a local scale height $z_0 = (\sqrt{6}\,\mu_0)^{-1}$ = 314 pc. This is consistent with literature values[21] of $\rho_0 = 0.097^{+0.013}_{-0.013}$ M$_\odot$pc$^{-3}$ and $z_0 = 280^{+70}_{-50}$ pc. We rerun our analysis using two additional models[12,13] of the Galactic potential including their respective adopted $r_\odot$ values (so in total the same three Milky Way potentials applied to model the Wave's in-plane motion), finding no effect on our results.

By assuming vertical energy conservation, the period *T* with which each part of the Radcliffe Wave oscillates through the Galactic disk can be computed by:

$$\mathrm{T}\left(z_{max}, r\right) = 2\sqrt{2} \int_0^{z_{max}} \left(\Phi\left(z_{max}, r\right) - \Phi(z, r)\right)^{-0.5} dz \qquad (2)$$

where $z_{max}$ is the maximum deflection a subregion of the Wave can reach above the Galactic plane. Inserting equation (1) in equation (2) using the definition of the frequency of vertical oscillation $\omega(z_{max}, r) = 2\pi\, T^{-1}(z_{max}, r)$ and introducing the elliptic integral of first kind K, we can derive a fully analytic expression of the oscillation frequency:

$$\omega\left(z_{max}, r\right) = \frac{\pi}{2}\, \omega_0\, \exp\left((r_\odot - r)\left(2\,r_0\right)^{-1}\right) \left(1 - \mu_0^2\, z_{max}^2\right)^{0.5}\, \mathrm{K}^{-1}\left(\mu_0\, z_{max}\left(1 - \mu_0^2\, z_{max}^2\right)^{-0.5}\right). \qquad (3)$$

Introducing the distance $s$ along the Wave (projected onto the x-y plane) such that $s = 0$ pc at the beginning of the Wave (near CMa OB1), we note that based on the position of the Wave in the Galactic plane (pitch angle ≠ 0°) the Galactocentric radius $r$ can be parametrized as a function of this distance $s$ along the Wave, implying $\omega(z_{max}(s), r(s)) = \omega(s)$.

Note that, consistent with the literature,[14] we treat the underlying Gravitational potential as stationary. This choice is furthermore motivated by the distribution of the old stellar component around the Radcliffe Wave, which is expected to dominate the local Galactic potential.[14] In the old stellar component, there is no evidence[2,3] for perturbations with wavelength < 10 kpc. In the vicinity of the Radcliffe Wave, a perturbation of the potential with wavelength > 10 kpc would correspond to a local tilt in the potential. The Galactic warp also constitutes a large-scale change in the potential. However, we already examined the inclusion of a tilt and found that including two extra angle parameters has no effect on our results. In addition, perturbations with wavelength > 10 kpc, which can also be seen as a warping of the Galactic disk, are expected to change on timescales larger than the period of the Radcliffe Wave.[46]

There is also recent evidence for radially evolving vertical waves near the Sun[47,48] with wavelengths < 10 kpc. Since these waves appear radially (pitch angle close to 90°) while the Radcliffe Wave evolves predominantly tangentially (pitch angle close to 0°) in the Milky Way, these waves should not significantly affect the Radcliffe Wave.

The original model for the Radcliffe Wave[1] was purely *spatial*. Now, as we have both *spatial* and *kinematic* information of the Wave, we apply a new model to describe the overall spatial and kinematic behavior of the Radcliffe Wave. Although our new model has fewer free parameters than the original one,[1] it provides a better fit to the 3D spatial data (see details in "Modeling the Spatial Molecular Cloud Distribution"). In the new framework, the Radcliffe Wave's undulation is modeled by a damped sinusoidal wave perpendicular to the plane of the Milky Way. We use a quadratic function in $x$ and $y$ (Heliocentric Galactic Cartesian coordinates) described by three anchor points $(x_0, y_0)$, $(x_1, y_1)$, and $(x_2, y_2)$ to model the slightly curved baseline of the structure in the Galactic plane, as a linear model is insufficient to account for the observed Radcliffe Wave's curvature in the x-y plane. We examine the effect of adding in additional parameters used to account for an inclination of the structure with respect to the Galactic disk in the original (spatial) model for the Radcliffe Wave.[1] However, we find the results to be fully consistent with the structure oscillating through the x-y plane, so these additional parameters describing the inclination are not included in the final analysis.

We model the Wave as two counter-propagating waves with frequencies of ±ω but different amplitudes, in order to allow motion in the form of a traveling wave, a standing wave, or intermediate cases. In this context, we introduce a parameter $B$ which determines the mixture between a standing and traveling wave model, leading to the following equation:

$$z(s,t) = \zeta(s)\left(B \sin\left(\omega_0 t + 2\pi \Lambda(s) + \varphi\right) + (1-B) \sin\left(-\omega_0 t + 2\pi \Lambda(s) + \varphi\right)\right) \quad (4)$$

$$v_z(s,t) = \zeta(s)\,\omega(s)\left(B \cos\left(\omega_0 t + 2\pi \Lambda(s) + \varphi\right) - (1-B) \cos\left(-\omega_0 t + 2\pi \Lambda(s) + \varphi\right)\right) \quad (5)$$

where $t$ denotes the time, $s$ the distance along the Wave, $\omega$ is defined by equation (3), $\omega_0$ is the zeroth order oscillation frequency, $\varphi$ is the phase and $\Lambda(s)$ and $\zeta(s)$ describe damping functions along the Wave.

In principle, $\Lambda(s)$ is a function of the time $t$. The dependence of $t$ results from the second and higher order terms in equation (3) which we denote as $|f(s)\omega_0|$. Using our adopted Gravitational potential, we compute $f$ from equation (3) to show that $f$ becomes at most ~0.25 for any distance $s$ along the Wave. Ultimately, we find $|\omega_0 t| \sim 0.4$, implying that $|f(s)\omega_0 t|$ is of the order ~0.03π. This allows us to neglect the dependence of time $t$ and approximate $\Lambda(s)$ as a pure function of $s$. Note that the time-independent approximation breaks down as soon as the oscillation proceeds into the future, implying that we can no longer set an upper limit on $|\omega_0 t|$. Deviations from the time-independent approximation would be comparable to the noise level after roughly 40 Myr. Therefore, we let the data in Supplementary Figure 5 fade out with time. We evolve the model with the horizontal damping time independent, in order to demonstrate the key kinematic feature of the Wave, i.e. that its kinematic behavior is consistent with a traveling wave.

To characterize the horizontal damping of the Wave, we define the phase $\phi_k = \arctan(z_k\, \omega_k\, v_{z,k}^{-1})$ of a cluster $k$ along the Wave. Ultimately, we find that $B$ is close to 1. Nevertheless, let us assume $B = 0.5$, meaning that the Wave corresponds to a standing wave. From equations (4) and (5) we conclude that the phases $\phi_k$ should be roughly constant (varying at most on the order of $|f(s)\omega_0 t| \sim 0.03\pi$) as a function of distances $s_k$ along the Wave (where $s_k$ is the distance along the Wave $s$ for a cluster $k$). In contrast, we find that the phases $\phi_k$ change over the whole range of $-\pi/2$ to $\pi/2$, such that $B = 0.5$ is inconsistent with the data. We therefore concentrate on $B \sim 1$ from this point forth.

For the case $B \sim 1$ (corresponding to a traveling wave), we conclude from equations (4) and (5) that the phase $\phi(s)$ should change according to $\Lambda(s) + \omega_0 t + \varphi$. By examining the overall distribution of stellar clusters in position and velocity space (see Figure 1), we can conclude that the horizontal damping function as a function of $s$, i.e. $\Lambda(s)$, is a monotonically increasing and convex function of the distance along the Wave $s$. Since the arctan function only projects to phases between $-\pi/2$ to $\pi/2$, we shift each $\phi_k$ by $n_k\pi$ (with $n_k$ being an integer for a given cluster $k$) such that the change of phases $\phi_k$ with distance $s_k$ is consistent with the fact that change in phase is a monotonically increasing and convex function of the distance along the Wave. We find that $\Lambda(s)$ described by

$$\Lambda(s) = \frac{s}{p - \gamma s} \quad (6)$$

where $s$ denotes the distance along the Wave, $p$ the period of the Wave and $\gamma$ sets the rate of the period decay, is able to explain the observed change in phases $\phi_k$, while keeping the number of free parameters low. Given the change of $\Lambda(s)$ with the distance along the Wave $s$, we use the following equation to compute the change in the Wave's wavelength $\lambda(s)$:

$$\lambda(s)^{-1} = \left(\partial_{s'} \Lambda(s')\right)\bigg|_{s'=s} = p\,(p - \gamma s)^{-2} \,. \tag{7}$$

We could in principle take the parabolic model introduced in the original work[1] to fit the data using optimized values for the free parameters given our new kinematic constraints. However, in this case the first order period and the second order scaling factor both go to infinity, meaning that the first order term in the model vanishes and the wavelength $\lambda(s)$ goes to infinity for $s$ going to 0. Since for $\Lambda(s)$ described by equation (6) the wavelength $\lambda(s)$ (see equation (7)) is well defined for $s = 0$ pc, we adopt equation (6) for the horizontal damping instead. However, the distribution of the phases allows for a variation of the model. With the advent of even more precise data, future work should be able to better constrain how the wavelength changes with distance along the Wave.

Next, we motivate our choice of the vertical damping function $\zeta(s)$. Using the vertical gravitational potential introduced in this work, we calculate the vertical energy of the Wave and from that the maximal vertical deflection $z_{max}$ of a single star cluster $k$ under the vertical pull of the gravitational potential given its vertical position and vertical velocity. Following the equations (1), (4) and (5), we approximate $z_{max}(s)$ as follows:

$$z_{max}(s) \approx |\zeta(s)|\,\left(1 + 4\cos^2\left(2\pi\,\Lambda(s) + \varphi\right)\left(B^2 - B\right)\right)^{0.5} \,. \tag{8}$$

Ultimately, we find that $B$ is close to 1 and in particular $B > 0.5$. This implies that the second part in equation (8) will always be larger than zero and can be seen as a scatter term perturbing $|\zeta(s)|$. In other words, the overall behavior of $z_{max}(s)$ is captured in $|\zeta(s)|$. In particular, in the case of $B = 1$ (corresponding to a traveling wave) $z_{max}(s)$ is equal to $|\zeta(s)|$.

We find that a Lorentzian describes the behavior of $z_{max}(s)$ marginally better than a Gaussian which is why we adopt the former for $\zeta(s)$ compared to the original study.[1] However, especially around the peak region, the data is in favor of the Gaussian model. We conclude that a mixture model with an additional free parameter (e.g. realized by a Voigt-profile) is likely the true underlying damping. Since our data is too limited to constrain an extra free parameter properly, we keep the Lorentzian damping profile. However, the evidence in favor of a mixture model might imply that the Radcliffe Wave originally originated from two distinct structures, whose original damping profiles might still be imprinted on the data. The final form of $\zeta(s)$ can be described by the following equation:

$$\zeta(s) = -A\left(1 + \left(\frac{s - s_0}{\delta}\right)^2\right)^{-1} \tag{9}$$

where $A$ denotes the amplitude, $s_0$ the position of the maximal deflection, and $\delta$ sets the rate of the amplitude decay.

As mentioned earlier, the vertical motion is decoupled from the Radcliffe Wave's in-plane motion. Therefore, the entire model is described by 16 parameters, which can be separated into two uncorrelated subgroups $\boldsymbol{\Theta}_1 = (x_0, y_0, x_1, y_1, x_2, y_2, A, p, \varphi, s_0, \delta, \gamma, \omega_0 t, B)$ and $\boldsymbol{\Theta}_2 = (v_{x,\text{plane}}, v_{y,\text{plane}})$.

In the following, we apply our overall model for the Radcliffe Wave oscillation to: just the molecular clouds as traced by 3D dust-mapping (as in the original study[1]), just the star clusters, and then the clouds together with the star clusters. We find that all three fits using different observational components and their combinations lead to consistent results, confirming that the Radcliffe Wave is a spatially and kinematically coherent structure composed of molecular clouds and young stellar clusters. Ultimately, we investigate the Wave's in-plane motion using the kinematic stellar information.

**Modeling the Spatial Molecular Cloud Distribution**

As in the original spatial-only fitting,[1] in this section we apply our overall wave model to the Radcliffe Wave's dust component, described by the corresponding 3D cloud positions from the *Major Cloud Catalog*[29,30] and the *Tenuous Connections Catalog.*[1] We include in our analysis only those molecular clouds for which we can obtain a radial velocity measurement (see "Line-of-sight Gas Velocities" section, below).

Note that for the cloud component we only have spatial information. In the absence of any kinematic constraint, we can always set $B$ and $\omega_0 t$ to fixed values, since equation (4) can be formulated as $\zeta(s)\,\eta\,\sin(2\pi\Lambda(s) + \varphi + \kappa)$ where $\eta$ and $\kappa$ are functions of $B$ and $\omega_0 t$. We choose $B = 1$ and $\omega_0 t = 0$, leading to $\eta = 1$ and $\kappa = 0$.

To model the data, we define the spatial distance vector $\mathbf{d}_{\text{cloud},i}$ of the i-th molecular cloud of the Radcliffe Wave relative to the wave model as

$$\mathbf{d}_{\text{cloud},i}(s) = (x_i, y_i, z_i) - (x(s), y(s), z(s)) \tag{10}$$

where $s$ is the distance along the Wave in the Galactic plane. The original sampling of the Radcliffe Wave[1] was not chosen to provide equal coverage over the entire structure, so some regions (e.g. Orion) have much higher sampling than other regions (e.g. North American Nebula). To consider each part of the Radcliffe Wave with equal importance, we introduce weights $w_{\text{cloud},i}$ derived using a kernel density estimation approach with a Gaussian kernel whose bandwidth is chosen to be a few percent of the total length of the Wave. Given our definition of the log-likelihood, the weights are set to sum to the number of molecular clouds in our sample. Assuming that the positions of the molecular clouds have been derived independently, the log-likelihood for a given realization of our model is

$$\ln\left(L\left(\boldsymbol{\theta}_1\right)\right) = -\sum_i w_{\text{cloud},i} \left( \frac{\mathbf{d}^2_{\text{cloud},i}(s_{min})}{2\,\sigma^2} + \ln\left(2\pi\,\sigma^2\right) \right) \tag{11}$$

where $s_{min}$ is defined as the distance along the Wave which minimizes $\mathbf{d}^2_{\text{cloud},i}$ for a given cloud $i$ and the corresponding distance vector from equation (10). In addition, we introduced a scatter parameter $\sigma$ which indicates the spread of the clouds around the model. Assuming that the relative cloud-model distance-vectors are normally distributed, $d^2_{min} = \Sigma_i\,\mathbf{d}^2_{\text{cloud},i}(s_{min})$ is chi-squared distributed with two degrees of freedom such that $\sigma^2$ is calculated as 0.5 times the standard deviation of $d^2_{min}$. Inferring the values of $\boldsymbol{\Theta}_1$ in a Bayesian framework, we sample for our free parameters using the nested sampling code *dynesty*.[49] Based on initial fits, we adopt pairwise independent priors on each parameter which are described in Extended Data Table 1 and run a combination of random walk sampling with multi-ellipsoid decompositions and 1000 live points. The results of our sampling procedure are summarized in Extended Data Table 2. We find a radius of the Wave of 47 pc defined by the Wave's scatter parameter $\sigma$. By calculating the weighted and unweighted sum of the squared cloud model distances for the new and the original[1] wave model using equation (10), we find in both cases (weighted and unweighted) that the ratio of the sums between the new and the original model is less than one, implying that the new model, in spite of having fewer free parameters, actually fits the 3D spatial molecular cloud data better.

**Modeling the Oscillation of the Wave's Stellar Component**

In this section we apply our overall wave model to the Radcliffe Wave's stellar component, as traced by star clusters. To fit our model to the Wave's star clusters, we define the spatial and kinematic distance vector $\mathbf{d}_{\text{cluster},k}$ of the $k$-th cluster relative to the wave model as

$$\mathbf{d}_{\text{cluster},k}(s) = \left(x_k, y_k, z_k, v_{z,k}\,\omega_0^{-1}\right) - \left(x(s), y(s), z(s), v_z(s)\,\omega_0^{-1}\right) \, . \tag{12}$$

To consider positions and velocities with equal importance, we multiplied the fourth entry of $\mathbf{d}_{\text{cluster},k}$ by $\omega_0^{-1}$. Assuming that the positions and velocities of the clusters have been derived independently, the log-likelihood in this case is given by

$$\ln\left(L(\boldsymbol{\Theta}_1)\right) = -\frac{1}{2}\sum_k w_{\text{cluster},k}\left(C\left(\frac{\mathbf{d}_{\text{cluster},k}(s_{min})}{\sigma}\right)^2 + 3\ln(2\pi\sigma^2)\right) \tag{13}$$

where $s_{min}$ is defined as the distance along the Wave which minimizes $\mathbf{d}^2_{\text{cluster},k}$ for a given cluster $k$ and the corresponding distance vector from equation (12). The scatter parameter $\sigma$ remains the same as in the molecular cloud discussion, and, to again consider each part of the Radcliffe Wave with equal importance, weights $w_{\text{cluster},k}$ were implemented, derived using a kernel density estimation approach with a Gaussian kernel whose bandwidth is chosen to be a few percent of the total length of the Wave. Given our definition of the log-likelihood, the weights are set to sum to the number of clusters in our sample. In addition, to search for outliers, we introduced an outlier model represented by a truncation function $C$. Note that in the case of the Radcliffe Wave's molecular clouds no outlier model was applied as the selection used for the dust already included outlier modeling in the original study.[1] We choose $C(x_k) = (\Sigma_n c_{k,n})^{0.5}$ with $c_{k,n} := \tanh(x^2_{k,n})$ where $x_{k,n}$ describes the n-th component of $x_k$ ($n$ going from 1 to 4), satisfying that $C(x_k) = \|x_k\|$ holds for small $\|x_k\|$, and that clusters with large distances between observation and model have little effect on the log-likelihood.

To allow for a traveling wave, a standing wave, and intermediate cases we apply a uniform prior from 0 to 1 for the parameter $B$ (see Extended Data Table 1). We infer the values of $\boldsymbol{\Theta}_1$ in a Bayesian framework, sampling for our free parameters using the nested sampling code *dynesty*. We adopt pairwise independent priors on each parameter (see Extended Data Table 1), and run a combination of random walk sampling with multi-ellipsoid decompositions and 1000 live points. The results of our sampling procedure are summarized in Extended Data Table 2.

We find a best fit value for $B$ = 0.84. The value of $B$ closer to 1 indicates that data favors a traveling wave. In addition, we redo the same fit with the same free parameters $\boldsymbol{\Theta}_1$ while fixing $B$ = 1 and $\omega_0 t$ = 0. By comparing the logarithm of the Bayes factor of the two realizations ($B$ = 1 and $B$ varying between 0 and 1) given their evidences $Z_1$ and $Z_2$, we find $\Delta \ln(Z) = \ln(Z_2) - \ln(Z_1) \approx 1$. Thus, both models are an equivalently good fit to the data, while the realization with fixed B = 1 is slightly preferred. We therefore conclude that the Wave's motion is of first order consistent with a traveling wave, implying that we can set $B$ = 1 in all further performed calculations, such that $\boldsymbol{\Theta}_1$ is reduced by 2 parameters, implying $\boldsymbol{\Theta}_1 = (x_0, y_0, x_1, y_1, x_2, y_2, A, p, \varphi, s_0, \delta, \gamma)$.

To investigate the effect of the Sun's current vertical position with respect to the Galactic plane, we rerun our analysis including the vertical position $z_0$ of the Sun as a free parameter. We set a uniform prior with a lower limit of $z_0$ = -10 pc and upper limit $z_0$ = 25 pc. After marginalizing over $z_0$ we find that the amplitude $A$, the phase $\varphi$, and the damping parameter $\delta$ change by less than a third of the 1$\sigma$ uncertainty reported in the case where $z_0$ is fixed. All remaining parameters in $\boldsymbol{\Theta}_1$ change by less than 10% of the 1$\sigma$ uncertainty reported in the case where $z_0$ is fixed. Therefore, we perform all calculations with the Sun located in the plane, consistent with recent work in this regard.[50]

**Modeling the Oscillation of the Wave's Stellar and Molecular Cloud Component**

Finally, we model the spatial positions of clouds together with the stellar clusters positions and velocities. The combined log-likelihoods for both cases give the best-fit model shown in Figure 1 as well as its corresponding parameters summarized in Extended Data Table 2. In Figure 1, we encode the transparency of the young stellar clusters as opaque for statistical inliers and transparent for outliers. Based on the previous section, we define a cluster $k$ to be an outlier if for any value of $n$ $c_{k,n}$ is larger than tanh(3), meaning that we only include clusters which lie inside a 3$\sigma$-radius, where $\sigma$ is the Wave's scatter parameter. The majority of the clusters identified as outliers belong to the Perseus and Taurus molecular clouds. In particular, since all clusters associated with Perseus and Taurus are identified as outliers, we argue that Perseus and Taurus seem to be kinematically distinct from the Radcliffe Wave. Their kinematics is likely dominated by the expansion motions of the "Local" and "Perseus-Taurus" feedback bubbles[9,15] on whose surfaces the clouds are located. Most of the remaining outliers can be assigned to the Orion star-forming region. It is important to note, however, that the fraction of outliers compared to inliers in Orion is no more than a quarter, implying that it follows the general Radcliffe Wave motion, so only a fraction of Orion's clusters' motions is dominated by internal feedback.[51,52]

In Figure 1, the typical uncertainty in the positions and in the vertical velocity of the stellar clusters and the typical uncertainty in the positions of the molecular clouds are of the same order of magnitude as the size of the corresponding symbols. The method used to derive the cloud distances introduces an additional systematic cloud distance uncertainty of about 5%.[29,30] Note that the small ripples at the right end of the Wave naturally result from the linear damping model of the Wave's period with distance. Since we have no star cluster measurements at the right end of the Wave (around CygnusX) and the amplitude of the ripples in this region gets close to the error of the dust distance measurements,[29,30] we can not distinguish between ripples and no ripples based on the data, indicated by the gray color of the fit in Figure 1. This does not affect the overall Wave's best-fit or the obtained properties of the Wave, as our results are dominated by the data with larger deflections above and below the Galactic plane.

Using the best-fit parameters and equation (7), we find a mean wavelength of around 2000 pc, with an underlying range of wavelengths varying from around 400 pc (near Cygnus X) to 4000 pc (near CMa OB1). Given the change in wavelength and the vertical damping of the Wave, we calculate its phase velocity using the ratio between the wavelength $\lambda(s)$ (see equation (7)) and the period T($s$) (see equation (2)) at any distance $s$ along the Wave. We find that the phase velocity changes from ~40 km s$^{-1}$ (near CMa OB1) to ~5 km s$^{-1}$ (near Cygnus X). We explore whether a variation in the local matter density may explain this order-of-magnitude change in the phase velocity. We find that the local matter density $\rho_0$ would need to change by at least two orders of magnitude, which is inconsistent with observations or models of the Milky Way.[14,21] However, this change in phase velocity may provide clues to the Wave's origin. Since, in addition to the large change in the phase velocity, a two-component damping function provides the best fit to the vertical damping of the Wave, we hypothesize that the Radcliffe Wave may have originated from the interplay of two different physical processes. Future simulations should shed light on whether such a combination could produce structures on the right scale that are also kinematically consistent with a traveling wave.

**Modeling the Wave's In-plane Motion**

In this section, we investigate the in-plane motion of the Radcliffe Wave described by the values of $\boldsymbol{\Theta}_2$ = ($v_{x,plane}$, $v_{y,plane}$). For this purpose, we subtract Galactic and differential rotation using an existing model of the Galactic potential[45] including its updated parameters[11] from the observational values of the $x$- and $y$-velocity ($v_{x,k}$, $v_{y,k}$) of the $k$-th cluster. We tested two alternative models for the Galactic potential,[12,13] finding no effect on our results. We apply a standard chi-square minimization with independent priors on each $v_{x,plane}$ and $v_{y,plane}$ (see Extended Data Table 1) to search for the best values to describe the observed in-plane motion. For the chi-square minimization, we adopt a typical error of 2 km s$^{-1}$, derived by taking the median $v_x$ and $v_y$ error across the sample. By computing the 16th, 50th and 84th percentiles of the samples we find $v_{x,\,plane}$= -3.9 $^{+2.6}_{-2.7}$ km s$^{-1}$ and $v_{y,\,plane}$ = 3.1 $^{+2.7}_{-2.6}$ km s$^{-1}$. This corresponds to a motion of 4.9 km s$^{-1}$ radially away from the Solar system as well as of 1.0 km s$^{-1}$ parallel to the Radcliffe Wave, implying that the Radcliffe Wave is radially drifting away from the Galactic Center with a velocity of around 5 km s$^{-1}$. In Extended Data Figure 2 we show the Wave's radial and tangential velocity vectors as well as the underlying stellar cluster data used to derive the kinematic properties. In Panel a and c of Extended Data Figure 2, the Galactic coordinate $x$-$y$ frame is rotated anticlockwise by 62°. The $x$-axis in this rotated frame is parallel to an axis along the Radcliffe Wave and will be referred to as "position along the Wave," throughout this work. The direction of this motion suggests that in the past, the Radcliffe Wave may have been at the same location where the star clusters Upper Centaurus Lupus and Lower Centaurus Crux were born 15-16 Myr ago,[9] potentially providing the reservoir of molecular gas needed for their formation.

**Line-of-sight Gas Velocities**

In this section, we consider the dynamics of the Wave using line-of-sight velocity measurements[8] of $^{12}$CO, a tracer of gas motion independent from that used in our 3D dust and clusters best-fit analysis described above. One may wonder how modeling 3D motion is possible when spectral line measurements only probe a 1D line-of-sight velocity, $v_{LSR}$. Since the Wave is so closeby (250 pc at closest), so long (~3 kpc), and extends so far above the Galactic disk (~200 pc), various lines of sight to its clouds actually sample a range of combinations of $v_x$, $v_y$, and $v_z$. Extended Data Figure 3b and the interactive Supplementary Figure 3 illustrate this lucky circumstance with yellow lines-of-sight to the Wave that a viewer will note are inclined at a wide variety of angles relative to an *x-y-z* frame.

We measure line-of-sight cloud velocities ($v_{LSR}$) using a Galaxy-wide $^{12}$CO survey[8] with an angular resolution of 0.125 deg and LSR velocity resolution of 1.3 km s$^{-1}$. We exclude regions far beyond the Radcliffe Wave by applying an LSR velocity cutoff of 30 km s$^{-1}$ . For each set of spatial cloud coordinates, we compute a $^{12}$CO spectrum over a region with a radius of 0.3 deg at the corresponding position on the sky. A radial velocity is then assigned to each spatial position using non-linear least squares fitting of Gaussians based on the Levenberg-Marquardt algorithm.[53] For each Gaussian, we assign the mean as the radial velocity and the 1$\sigma$-deviation as the corresponding error. Nine percent of the clouds fall outside the limits of the $^{12}$CO survey or show no emission above the noise threshold, so we exclude these clouds from our phase space analysis. In Extended Data Figure 3a we show the results of the spectral-fitting alongside our best model for the motion of the Radcliffe Wave. The computation of the errors shown in Extended Data Figure 3a involved combining multiple measurements of a single molecular cloud into a single data point. To this end, we average over each single cloud and present the obtained 16th and 84th percentiles. The comparison between data and model in this figure is based on the projection of our 6D (*x, y ,z ,$v_x$ ,$v_y$ ,$v_z$*) best-fit model onto the 4D phase space (*x, y, z, $v_{LSR}$*) of the gas data. The theoretical description of the radial velocity $v_{\text{rad}}$ is derived using

$$v_{LSR,i}(\mathbf{v}_i) = \cos(l_i)\cos(b_i)v_{x,i} + \sin(l_i)\cos(b_i)v_{y,i} + \sin(b_i)v_{z,i} = \mathbf{x}_i \cdot \mathbf{v}_i \qquad (14)$$

where $l_i$ and $b_i$ are Galactic longitude and latitude of the *i*-th molecular cloud, $\mathbf{x}_i$ denotes its normalized position in Galactic Cartesian coordinates and $\mathbf{v}_i$ the corresponding velocity with components $v_{x,i}$, $v_{y,i}$ and $v_{z,i}$. We show in Extended Data Figure 3a that our model explains the $^{12}$CO observations, meaning that the velocities of the molecular clouds, for which we can only study an incomplete phase space, match the motion of the young stellar clusters. As a consequence, we can transfer our results from the full phase-space study of the stars to all components of the Radcliffe Wave, including the gas, and characterize its full structure and kinematics.

**The Sun's Vertical Oscillation Period and Local Properties of the Galactic Potential**

In this section we present a first application of the Radcliffe Wave's oscillation, using it to measure the Sun's vertical oscillation frequency and local properties of the Galactic potential.

The Radcliffe Wave shows a systematic pattern in its gas and star cluster velocities. Above, we show how that pattern is consistent with oscillation in the Galaxy's gravitational potential. But, we can also use that pattern – without prior knowledge about the vertical gravitational potential – to infer a maximum amplitude as well as a maximum velocity for each star forming region along the Wave. As a consequence, we can investigate the vertical oscillation frequency without needing to adopt a standard model of the Milky Way gravitational potential.[11–13] We can essentially reverse the setup of our study and directly measure the oscillation frequency described by equation (3), using only the observed Radcliffe Wave oscillation. For this purpose, we fit our model for the coherent oscillation of the Radcliffe Wave, described by equations (4) and (5) to the observations by treating the variables $\omega_0$, $\mu_0$ and $r_0$, which define the oscillation frequency calculated in equation (3), as free parameters. Earlier, these parameters were determined by comparing the Galactic potential described by equation (1) with a full model of the Galactic potential to show that the Radcliffe Wave oscillation is consistent with the current understanding of the Galactic potential of the Milky Way.[11–13] The entire model is now described by 15 parameters $\Theta_3 = (x_0, y_0, x_1, y_1, x_2, y_2, A, p, \varphi, s_0, \delta, \gamma, \omega_0, \mu_0, r_0)$.

Inferring the values of $\Theta_3$ in a Bayesian framework, we sample for our free parameters using the nested sampling code *dynesty*[49] and, based on initial fits, we adopt pairwise independent priors on each parameter which are described in Extended Data Table 1. We run a combination of random walk sampling with multi-ellipsoid decompositions and 1000 live points. The results of our sampling procedure are summarized in Extended Data Table 2. We find a median value and $1\sigma$ errors (computed using the 16th, 50th and 84th percentiles of the samples) of $\omega_0 = 68^{+8}_{-7}$ km s$^{-1}$ kpc$^{-1}$, $\mu_0 = 1.4^{+0.2}_{-0.2}$ kpc$^{-1}$ and $r_0 = 3.6^{+0.5}_{-0.5}$ kpc. These values are consistent with the previously adopted values for $\omega_0$, $\mu_0$, and $r_0$ within the $1\sigma$ error. By comparing the logarithm of the Bayes factor[54] of the two realizations (fixed and varying oscillation frequency) using their evidences $Z_1$ and $Z_2$, we find $\Delta \ln(Z) = \ln(Z_2) - (Z_1) \approx 1$, meaning that both models are an equivalently good fit to the data[54].

Following the vertical density model of an isothermal sheet based on a Maxwellian vertical velocity distribution,[55] we calculate the parameters defining the Galactic mass distribution: the midplane density $\rho_0 = \omega_0^2 (4 \pi G)^{-1}$ and the vertical scaleheight $z_0 = (\sqrt{6} \mu_0)^{-1}$. This leads to $\rho_0 = 0.085^{+0.021}_{-0.017}$ M⊙pc$^{-3}$ and $z_0 = 294^{+63}_{-43}$ pc. To compute errors on $\rho_0$ and $z_0$, we sample from the probability distribution of $\omega_0$ and $\mu_0$ to obtain 2000 realizations of the overall probability distributions. From that, we compute the median value and $1\sigma$ errors using the 16th, 50th and 84th percentiles of the samples. Our results are consistent with the conventional approach of deriving properties of the local Galactic mass distribution.[21,24–26] However, we caution that our constraint on the scale height of $z_0 \sim 294$ pc is tempered by the fact that a scale height of $\sim 294$ pc leads to a 7% change of the total oscillation frequency. Given that we determine $\omega_0$ with a $\sim 10\%$ accuracy, we emphasize that for $z_0 > 300$ pc, the sensitivity of our model to $z_0$ would be indistinguishable from the noise level of this study.

The derived midplane density contains both the local amount of dark matter and the baryonic matter component. By considering observations of the baryonic mass distribution near the Sun[21] of $\rho_{0,baryons} = 0.084^{+0.012}_{-0.012}$ M⊙pc$^{-3}$ we infer the amount of local dark matter. Assuming a spherical dark halo, we obtain a local dark matter density of $\rho_{0, \text{dark halo}} = 0.001^{+0.024}_{-0.021}$ M⊙pc$^{-3}$, consistent with the conventional approach to determine the Galactic potential's properties at the $1\sigma$ level.[21,24–26] Moreover, we obtain a total surface density $\Sigma_0 = 2 z_0 \rho_0$ of 50.3 $M_\odot pc^{-2}$.

Based on equations (2) and (3) we additionally find the Sun's oscillation period to be $95^{+12}_{-10}$ Myr where we assumed a vertical velocity of the Sun[10] of 7.8 km s$^{-1}$. To compute the period's error bars, we use the probability distribution of each free parameter reflecting their measurement uncertainties and then randomly sample from the overall multivariate probability distribution 2000 times to create 2000 mock realizations. From that, we compute the median value and $1\sigma$ errors using the 16th, 50th and 84th percentiles of the samples. Again, we allow the current position of the Sun to vary from -10 to 25 pc below or above the Galactic plane. We find that the effect resulting from the Sun's vertical position is smaller than 0.5 Myr which is more than an order of magnitude smaller than the observed $1\sigma$ deviation. Therefore, all computations were done with the Sun lying in the plane. The maximal deflection $z_{max}$ of the Sun above the plane used in equation (3) is computed by assuming vertical energy conservation. The Galactocentric radius of the Solar system is chosen to be 8.4 kpc. We varied the Solar radius from 7.9 to 8.7 kpc, finding no effect on our results. This is expected as our analysis mainly depends on the distance between the Sun and the Radcliffe Wave which is well-constrained[1,29,30] and not on our exact position in the Milky Way.

We also infer the period of the Sun following up-to-date measurements of the baryonic matter component.[21] Using equation (2) with a potential corresponding to the vertical density model of an isothermal sheet, based on a Maxwellian vertical velocity distribution[55]

$$\Phi(z) = 4\pi G \rho_0 z_0^2 \ln\left(\cosh\left(z z_0^{-1}\right)\right) \tag{15}$$

with midplane density[21] $\rho_0 = 0.084$ $M_\odot pc^{-3}$ and the vertical scaleheight[21] $z_0$ defined via the surface density[21] $\Sigma_0 = 2 z_0 \rho_0 = 47.1$ $M_\odot pc^{-2}$, we compute the oscillation period which would follow purely from the visible matter as 95.3 Myr. To compare our method to the conventional approach of analyzing stellar dynamics by measuring the vertical gravitational acceleration,[23–26] we add to equation (15) the potential of a spherical dark halo which can be expressed in our region of interest as a harmonic oscillator[24]

$$\Phi(z) = 2\pi G \rho_0 z^2 \tag{16}$$

with a density[21] of $\rho_0 = 0.013$ $M_\odot pc^{-3}$. This leads to a total period of 88.2 Myr, consistent with our result within the $1\sigma$ error.

To account for a possible dark disk, we observe that the smaller the scale height of the dark disk, the smaller is the resulting oscillation period, while keeping the surface density of the dark disk constant. Therefore, we assume a comparatively[27,56] large vertical disk scale height of 25 pc and use this to derive an upper limit on the surface density that a hypothetical dark disk may have to be consistent with our measurement of the Solar oscillation period of 95 Myr. Inserting the sum of equation (15) with values for the baryonic mass component[21] and equation (15) with $z_0 = 25$ pc and $\Sigma_{0, dark\ disk}$ as free parameter in equation (2), we find an upper bound for the dark disk of $0.1^{+3.3}_{-3.3}$ $M_\odot pc^{-2}$. Note that this upper bound is constrained without any additional dark halo, which would only lower the upper bound further.

**Data Availability**

The datasets generated and/or analyzed during the current study are publicly available at the Harvard Dataverse.
The $^{12}$CO radial velocities for the Radcliffe Wave are available at https://doi.org/10.7910/DVN/Q7F4PC.
The stellar cluster catalog is available at https://doi.org/10.7910/DVN/XSCB9N.

**Code Availability**

The code used to derive the results is available from RK upon reasonable request. In this context, publicly available software packages, including *dynesty*[49] and *astropy*,[57] were used. The visualization, exploration and interpretation of data presented in this work were made possible using the *glue*[58] visualization software. The interactive figures were made possible by the *plot.ly python* library.

**Methods References**

## Acknowledgements

We thank Jon Carifio, Sarah Jeffreson, Eric Koch, Vadim Semenov, Gus Beane, Michael Rugel, Stefan Meingast, Andrew Saydjari, Josh Speagle, Chervin Laporte, Shmuel Bialy, Josefa Großschedl, Theo O'Neill, Lisa Randall, and Bob Benjamin for useful discussions. The visualization, exploration, and interpretation of data presented in this work were made possible using the glue visualization software, supported under NSF grant numbers OAC-1739657 and CDS&E:AAG-1908419. AG and CZ acknowledge support by NASA ADAP grant 80NSSC21K0634 "Knitting Together the Milky Way: An Integrated Model of the Galaxy's Stars, Gas, and Dust." CZ acknowledges that support for this work was provided by NASA through the NASA Hubble Fellowship grant #HST-HF2-51498.001 awarded by the Space Telescope Science Institute (STScI), which is operated by the Association of Universities for Research in Astronomy, Inc., for NASA, under contract NAS5-26555. JA was co-funded by the European Union (ERC, ISM-FLOW, 101055318). Views and opinions expressed are, however, those of the author(s) only and do not necessarily reflect those of the European Union or the European Research Council. Neither the European Union nor the granting authority can be held responsible for them. AB was supported by the Excellence Cluster ORIGINS which is funded by the Deutsche Forschungsgemeinschaft (DFG; German Research Foundation) under Germany's Excellence Strategy – EXC-2094-390783311.

## Author Contributions

RK led the work and wrote the majority of the text. All authors contributed to the text. RK, AAG, CZ and JA led the data analysis. RK, AAG, CZ, AB and JA led the interpretation of the results, aided by MF and CS. JA, MK and NMR led the compilation of the stellar cluster catalog. RK and CZ led the statistical modeling. RK and AB led the theoretical analysis. RK, AAG and CZ led the visualization efforts.

## Competing Interests

The authors declare that they have no competing financial interests.

## Author Information

Correspondence and requests for materials should be addressed to Ralf Konietzka
(e-mail: ralf.konietzka@cfa.harvard.edu).

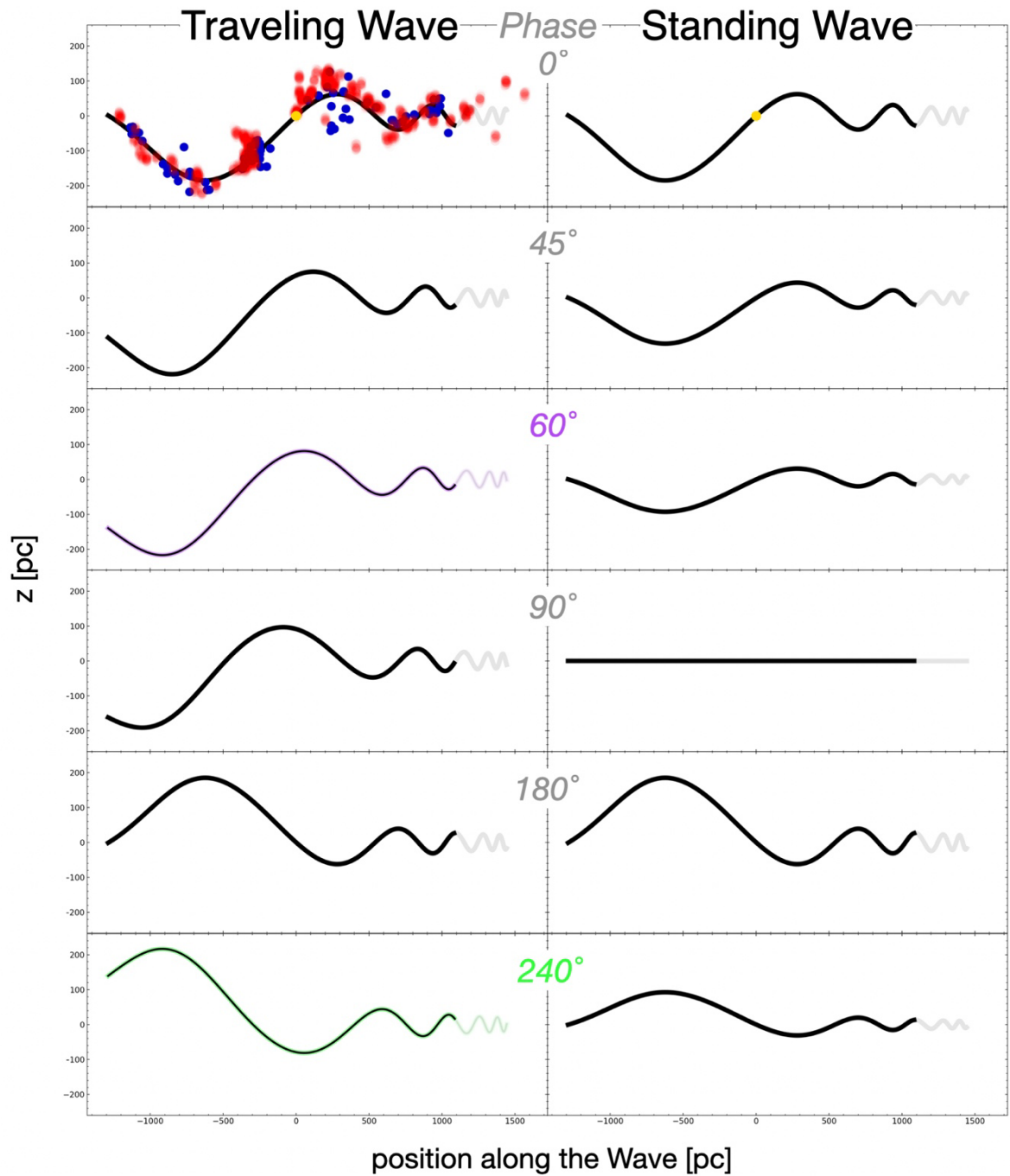

**Extended Data Figure 1.** *Comparison between a traveling and a standing wave. Selected phase snapshots are shown here. For the best experience, please view the online version in Supplementary Figure 4. In all Panels, the x-axis corresponds to the x-axis of a coordinate frame in which the Galactic coordinate xy-frame has been rotated anticlockwise by 62°.* **Left Panels:** *The evolution of the Radcliffe Wave (stellar cluster in blue, molecular clouds in red, the best-fit-model in black) with phase. The Sun is shown in yellow. The vertical motion the Radcliffe Wave is showing corresponds to a traveling wave. We show in purple (60°) and in green (240°) the same snapshots as in Figure 2.* **Right Panels:** *The evolution of the Radcliffe Wave starting from the best-fit-model in black if the Wave's motion corresponded to a standing wave.*

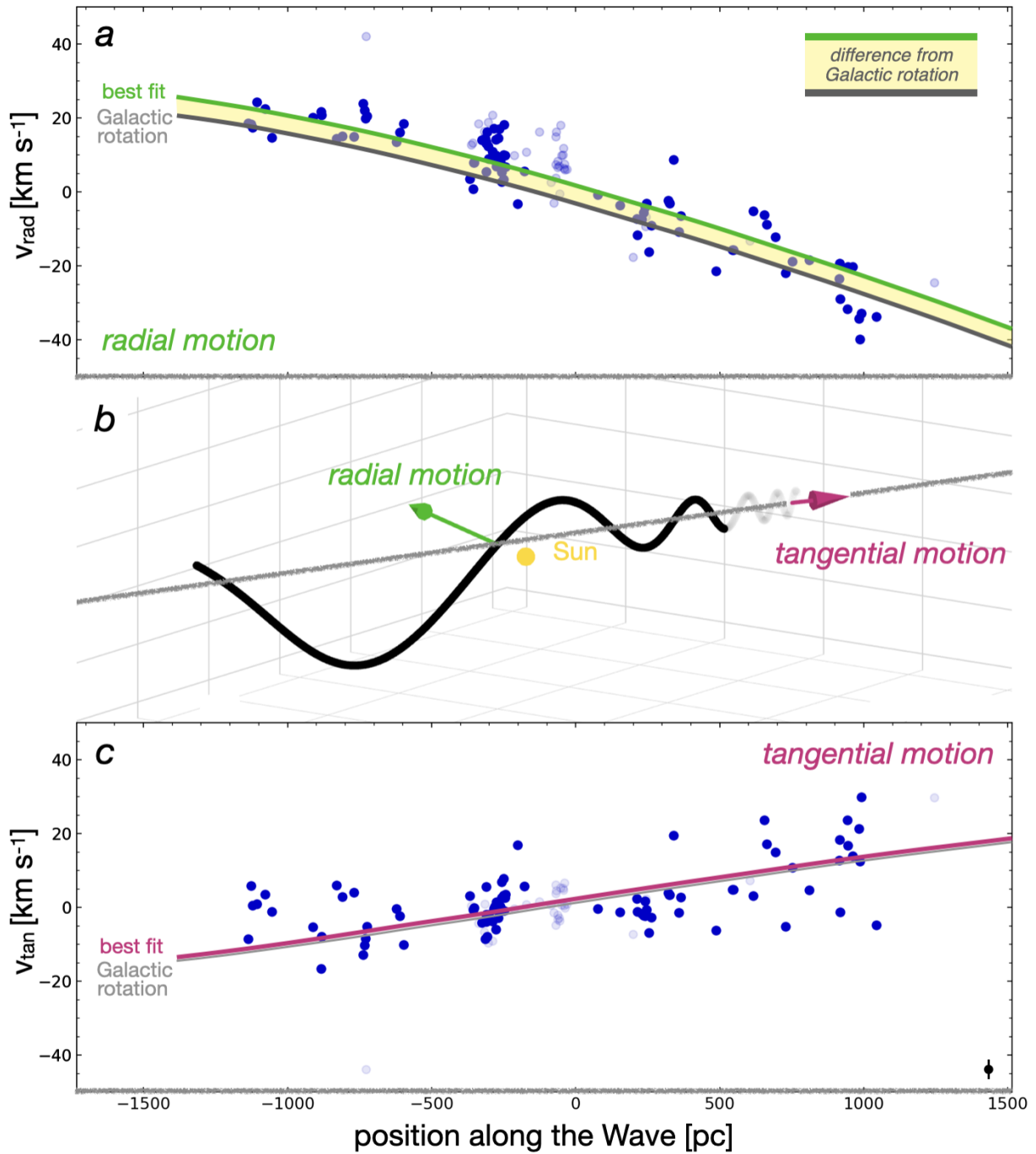

**Extended Data Figure 2:** *In-plane cluster velocities.* **Panel a**: *The radial in-plane cluster velocities are shown in blue (inliers opaque / outliers transparent). The best fit including Galactic rotation as well as the Wave's solid body motion of 4.87 km s$^{-1}$ in the radial direction (radially away from the Solar system, see Panel b) is shown in green. The x-axis corresponds to the x-axis of a coordinate frame in which the Galactic coordinate xy-frame has been rotated anticlockwise by 62°.* **Panel b (Interactive)**: *For the best experience, please view the online interactive version in* [Supplementary Figure 2](Supplementary Figure 2). *The best fit of the Radcliffe Wave is shown in black. The direction of the Wave's radial velocity (Panel a) is shown by the green vector. The direction of the Wave's tangential velocity (Panel c) is shown by the pink vector. The Sun is shown in yellow. The gray line corresponds to the x-axis of Panel a and c.* **Panel c:** *The tangential in-plane cluster velocities are shown in blue (inliers opaque / outliers transparent). The best fit including Galactic rotation as well as the Wave's solid body motion of 0.96 km s$^{-1}$ in the tangential direction (parallel to the Radcliffe Wave, see Panel b) is shown in pink. The effect of Galactic rotation is again shown in gray. The x-axis is the same as in Panel a. The typical uncertainty (1$\sigma$ errors) of the in-plane motion is shown in black in the right lower corner.*

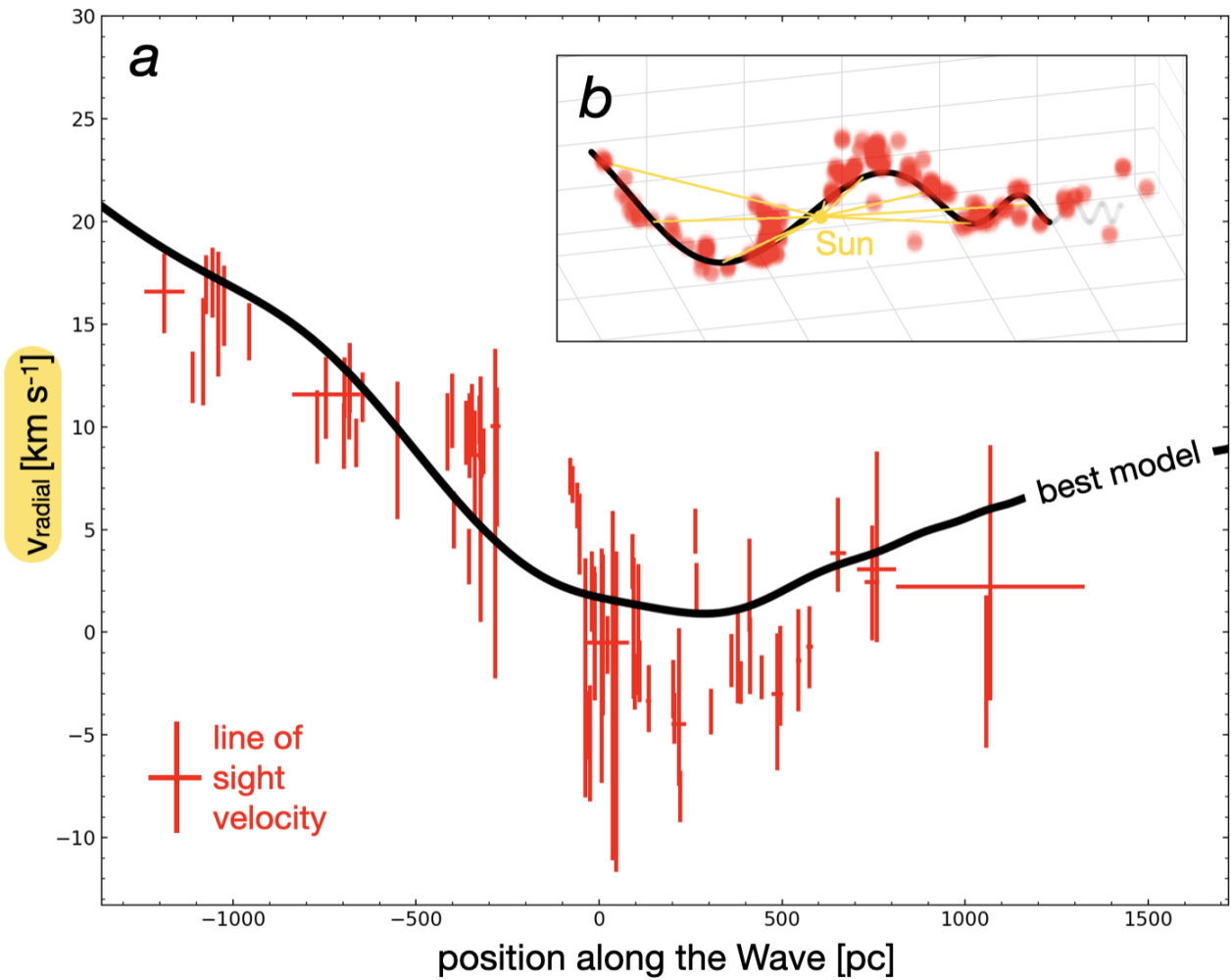

**Extended Data Figure 3:** *Radial $^{12}$CO Velocities.* **Panel a:** $^{12}$CO radial velocities of entire molecular clouds along the Radcliffe Wave derived from spectral-line fitting are shown in red, including their 16th and 84th percentiles (1$\sigma$ errors). Since we identified the Perseus and Taurus molecular clouds as outliers (see Methods), we do not show these clouds in this figure. The projection of our 6D best-fit model onto the 4D phase space of the gas data is shown in black. We find that our model explains the observed $^{12}$CO spectrum, which means the velocities of the molecular clouds, for which we can only study an incomplete phase space, match the motion of the young stellar clusters. The x-axis corresponds to the x-axis of a coordinate frame in which the Galactic coordinate xy-frame has been rotated anticlockwise by 62°. **Panel b (Interactive):** *For the best experience, please view the online interactive version in [Supplementary Figure 3](#). The Radcliffe Wave is shown spatially in 3D including the molecular clouds in red as well as the best fit in black. The Sun is shown with a yellow dot. The yellow lines represent a sample of lines of sight to illustrate the wide variety of angles along which the $^{12}$CO radial velocities were derived.*

| parameter | prior | parameter | prior |
| (1) | (2) | (3) | (4) |
| --- | --- | --- | --- |
| $x_0$ | $\mathcal{N}(-850\text{ pc}, 100\text{ pc})$ | $\ln(\gamma)$ | $\mathcal{N}(0.35, 0.07)$ |
| $y_0$ | $\mathcal{N}(-790\text{ pc}, 100\text{ pc})$ | $\varphi$ | $\mathcal{N}(-0.09\text{ rad}, 0.5\text{ rad})$ |
| $x_1$ | $\mathcal{N}(-255\text{ pc}, 100\text{ pc})$ | $\ln(\omega_0\text{ km}^{-1}\text{ s pc})$ | $\mathcal{N}(-2.4, 0.17)$ |
| $y_1$ | $\mathcal{N}(23\text{ pc}, 100\text{ pc})$ | $\ln(\mu_0\text{ pc})$ | $\mathcal{N}(-1.346, 0.25)$ |
| $x_2$ | $\mathcal{N}(313\text{ pc}, 100\text{ pc})$ | $r_0$ | $\mathcal{N}(3659\text{ pc}, 500\text{ pc})$ |
| $y_2$ | $\mathcal{N}(1368\text{ pc}, 100\text{ pc})$ | $\omega_0 t$ | $\mathcal{U}(-1.571\text{ rad}, 1.571\text{ rad})$ |
| $p$ | $\mathcal{N}(4650\text{ pc}, 300\text{ pc})$ | $B$ | $\mathcal{U}(0, 1)$ |
| $A$ | $\mathcal{N}(215\text{ pc}, 20\text{ pc})$ | $v_{x,\text{ inplane}}$ | $\mathcal{U}(-10\text{ km s}^{-1}, 10\text{ km s}^{-1})$ |
| $\ln(\delta\text{ pc}^{-1})$ | $\mathcal{N}(6.5, 0.1)$ | $v_{y,\text{ inplane}}$ | $\mathcal{U}(-10\text{ km s}^{-1}, 10\text{ km s}^{-1})$ |
| $s_0$ | $\mathcal{N}(680\text{ pc}, 200\text{ pc})$ | | |

**Extended Data Table 1**: *Priors on model parameters.* We set our priors on our parameters to be independent for each parameter, based on initial fits. $\mathcal{N}(\mu,\sigma)$ denotes a normal distribution with mean $\mu$ and standard deviation $\sigma$ and $\mathcal{U}(b_1,b_2)$ denotes a uniform distribution with lower bound $b_1$ and upper bound $b_2$.

| parameter / error | unit | molecular clouds | stellar cluster | stellar cluster + mixture model | molecular clouds + stellar cluster | molecular clouds + stellar cluster + changing oscillation period |
|---|---|---|---|---|---|---|
| (1) | (2) | (3) | (4) | (5) | (6) | (7) |
| $x_0$ | pc | -853.13 | -822.51 | -833.83 | -852.99 | -844.31 |
| $\pm \sigma_{x0}$ | pc | 51.12, -57.35 | 49.83, -51.82 | 49.57, -57.1 | 51.66, -54.91 | 49.13, -54.15 |
| $y_0$ | pc | -797.19 | -804.87 | -826.53 | -807.94 | -797.55 |
| $\pm \sigma_{y0}$ | pc | 48.74, -48.78 | 64.26, -60.12 | 63.65, -61.06 | 54.01, -50.29 | 49.32, -54.25 |
| $x_1$ | pc | -276.15 | -289.51 | -284.02 | -275.13 | -276.51 |
| $\pm \sigma_{x1}$ | pc | 45.72, -47.33 | 49.9, -48.11 | 50.64, -50.65 | 43.85, -46.1 | 42.69, -42.98 |
| $y_1$ | pc | 18.27 | 42.29 | 25.84 | 30.48 | 29.03 |
| $\pm \sigma_{y1}$ | pc | 89.0, -88.02 | 87.38, -85.05 | 88.24, -83.4 | 84.66, -86.94 | 81.8, -81.52 |
| $x_2$ | pc | 289.05 | 329.57 | 317.16 | 293.84 | 291.68 |
| $\pm \sigma_{x2}$ | pc | 28.05, -27.1 | 49.93, -46.41 | 51.18, -45.58 | 30.25, -26.96 | 32.72, -27.41 |
| $y_2$ | pc | 1386.24 | 1365.38 | 1377.48 | 1387.37 | 1379.77 |
| $\pm \sigma_{y2}$ | pc | 80.67, -79.49 | 93.43, -89.69 | 91.07, -96.67 | 85.95, -75.5 | 91.24, -73.37 |
| $p$ | pc | 4773.99 | 4752.21 | 4715.14 | 4779.9 | 4749.05 |
| $\pm \sigma_p$ | pc | 189.15, -187.63 | 207.28, -198.8 | 216.07, -214.2 | 195.49, -192.26 | 191.12, -182.03 |
| $A$ | pc | 219.88 | 216.52 | 229.32 | 218.64 | 222.92 |
| $\pm \sigma_A$ | pc | 18.34, -16.24 | 11.54, -10.95 | 13.23, -12.18 | 13.81, -11.41 | 13.32, -12.5 |
| $\delta$ | pc | 797.78 | 624.25 | 649.14 | 740.01 | 738.57 |
| $\pm \sigma_\delta$ | pc | 67.28, -62.43 | 60.19, -55.96 | 61.95, -56.19 | 58.78, -55.14 | 59.33, -52.89 |
| $s_0$ | pc | 482.71 | 673.18 | 662.89 | 544.99 | 517.02 |
| $\pm \sigma_{s0}$ | pc | 113.81, -103.35 | 104.02, -103.13 | 112.13, -107.4 | 89.88, -100.53 | 99.91, -91.41 |
| $\gamma$ | | 1.49 | 1.4 | 1.4 | 1.46 | 1.46 |
| $\pm \sigma_\gamma$ | | 0.07, -0.07 | 0.09, -0.07 | 0.1, -0.08 | 0.08, -0.07 | 0.09, -0.07 |
| $\varphi$ | rad | -0.12 | -0.23 | 0.01 | -0.15 | -0.14 |
| $\pm \sigma_\varphi$ | rad | 0.13, -0.14 | 0.16, -0.16 | 0.25, -0.23 | 0.13, -0.13 | 0.13, -0.14 |
| $\omega_0 t$ | rad | | | -0.38 | | |
| $\pm \sigma_{\omega_0 t}$ | rad | | | 0.19, -0.17 | | |
| $B$ | | | | 0.84 | | |
| $\pm \sigma_B$ | | | | 0.07, -0.06 | | |
| $\omega_0$ | km s$^{-1}$ kpc$^{-1}$ | | | | | 67.713 |
| $\pm \sigma_{\omega_0}$ | km s$^{-1}$ kpc$^{-1}$ | | | | | 7.677, -7.111 |
| $\mu_0$ | kpc$^{-1}$ | | | | | 1.377 |
| $\pm \sigma_{\mu_0}$ | kpc$^{-1}$ | | | | | 0.249, -0.247 |
| $r_0$ | pc | | | | | 3563.32 |
| $\pm \sigma_{r0}$ | pc | | | | | 513.7, -500.43 |

**Extended Data Table 2**: *Best fit parameters*. Best fit parameters for different underlying datasets and realizations are shown. See Methods section for more details. The errors are computed using the 16th, 50th and 84th percentiles of their samples. **(1)** Model parameter. **(2)** Corresponding unit. **(3)** Best fit parameters obtained by applying our model only to the spatial molecular cloud observations. **(4)** Best fit parameters obtained by applying our model only to the spatial and kinematic stellar cluster observations. **(5)** Same approach as in (4) but including a mixture model to allow for both, a standing and traveling wave. **(6)** Best fit parameters obtained by applying our model to the spatial and kinematic molecular cloud and stellar cluster observations. **(7)** Same approach as in (5) but leaving the parameters describing the vertical oscillation frequency as free parameters.